# A dual mode adaptive basal-bolus advisor based on reinforcement learning

Qingnan Sun, Marko V. Jankovic, João Budzinski, Brett Moore, Peter Diem, Christoph Stettler, Stavroula G. Mougiakakou, *Member IEEE*

*Abstract*— **Self-monitoring of blood glucose (SMBG) and continuous glucose monitoring (CGM) are commonly used by type 1 diabetes (T1D) patients to measure glucose concentrations. The proposed adaptive basal-bolus algorithm (ABBA) supports inputs from either SMBG or CGM devices to provide personalised suggestions for the daily basal rate and prandial insulin doses on the basis of the patients' glucose level on the previous day. The ABBA is based on reinforcement learning (RL), a type of artificial intelligence, and was validated *in silico* with an FDA-accepted population of 100 adults under different realistic scenarios lasting three simulated months. The scenarios involve three main meals and one bedtime snack per day, along with different variabilities and uncertainties for insulin sensitivity, mealtime, carbohydrate amount, and glucose measurement time. The results indicate that the proposed approach achieves comparable performance with CGM or SMBG as input signals, without influencing the total daily insulin dose. The results are a promising indication that AI algorithmic approaches can provide personalised adaptive insulin optimisation and achieve glucose control - independently of the type of glucose monitoring technology.**

## I. INTRODUCTION

**M**OST cases of diabetes can be broadly classified as type 1, where no insulin is secreted due to destruction of the

Manuscript received August 30, 2018
This research was carried out within the framework of the MyTreat research and development project, supported by the Swiss Commission of Technology and Innovation (CTI) under Grant 18172.1 PFLS-LS.

Q. Sun is with the with the ARTORG Center for Biomedical Engineering Research, University of Bern, 3008 Bern, Switzerland (e-mail: qingnan.sun@artorg.unibe.ch).
M. Jankovic is with the ARTORG Center for Biomedical Engineering Research, University of Bern, 3008 Bern, Switzerland, and the Department of Emergency Medicine, Bern University Hospital "Inselspital", 3010 Bern, Switzerland (e-mail: marko.jankovic@artorg.unibe.ch).
J. Budzinski is with the Debiotech S.A., 1004 Lausanne, Switzerland (e-mail: j.budzinski@debiotech.com).
B. Moore is with the St. Mary's University, School of Science, Engineering, and Technology, San Antonio, Texas, United States of America (e-mail: brett.moore@gmail.com).
P. Diem is with the Insel Gruppe AG, Freiburgstrasse 18, 3010 Bern, Switzerland (email: peter.diem@insel.ch).
C. Stettler is with the Division of Diabetes, Endocrinology, Clinical Nutrition and Metabolism, Bern University Hospital "Inselspital", 3010 Bern, Switzerland (e-mail: christoph.stettler@insel.ch).
S. Mougiakakou* is with the ARTORG Center for Biomedical Engineering Research, University of Bern 3008 Bern Switzerland, and the Division of Diabetes, Endocrinology, Clinical Nutrition and Metabolism, Bern University Hospital "Inselspital" (e-mail: stavroula.mougiakakou@artorg.unibe.ch).

pancreatic beta cells, or type 2, where either the pancreas does not produce enough insulin or the body does not effectively use the insulin produced.

The main goal of diabetes management is to maintain glucose levels within a healthy range, and this objective may require glucose monitoring and insulin therapy via pumps or injections. Diabetes patients may use two different approaches for insulin management: devices for the self-monitoring of blood glucose (SMBG), which measure glucose with one drop of finger blood several times during the day, or continuous glucose monitoring (CGM) systems, which use a subcutaneous miniaturised sensor to measure glucose levels every few minutes. In the case of adults with diabetes type 1 using SMBG, it is recommended to test glucose levels at least four times a day, i.e. before each meal and before going to bed [1]. Relatively few diabetic patients use CGMs [2], although this is expected to increase worldwide in the near future.

Insulin pumps deliver basal and bolus insulin. Basal insulin maintains glucose concentration at consistent levels during periods of fasting, while bolus insulin compensates for the effects of meal intake. Basal insulin is usually adjusted by the attending physician after reviewing the patient's glucose records, while the bolus dose is calculated using a bolus advisor. Bolus advisors use simple algorithms to estimate the insulin dose on the basis of the carbohydrate (CHO) content of the meal, the current blood glucose concentration (usually driven by an SMBG device), the patient's personal settings (e.g. correction factor, insulin-to-carbohydrate ratio - CIR), and the insulin on board [3].

As insulin sensitivity changes during the day, CIR and basal rate (BR) should be updated over time. Since the daily activities of diabetic patients tend to be repetitive, e.g. with respect to meal timing, meal amount etc., Owen *et al.* [4] incorporated a Run-to-Run (R2R) algorithm into a controller, which led to a more advanced bolus advisor. The advisor updated the bolus daily, using two postprandial SMBG measurements at 60 min and 90 min after the start of the respective meal. The advisor was clinically evaluated and gave promising results [5]. It has been proposed that bolus insulin could be estimated from CGM data, as supported by case-based reasoning (CBR) and R2R [6], [7]. The clinical safety of the algorithm has been presented in a single-arm pilot study [8]. The method has been extended to adjust the BR [9]. The *in silico* results indicated that this approach is of potential value.

The adaptation of BR was initially investigated by Palerm *et*

*al.* [5], using an R2R approach similar to the one presented in [4] and including five properly timed SMBG measurements. More recently, Toffanin *et al.* [10] adjusted the daily basal therapy using a number of well-established clinical indices, e.g. as derived from CGM data, and an R2R algorithm. The algorithm performed well in an *in silico* diabetic population.

The adaptation of BR and/or CIR has also been proposed within the framework of an artificial pancreas (AP). The AP provides an autonomous option for controlled insulin treatment - by combining an insulin pump, a CGM, and a control algorithm. Proportional–integral–derivative controllers (PIDs), model predictive controllers (MPCs), and fuzzy logic (FL) methods have been traditionally employed for clinically validated APs [11]. The MPC algorithm may, for example, be tuned by employing an R2R approach. This adapts the BR during the night and the CIR during the day [10], [12]. An R2R approach, together with CBR, was used within a closed-loop controller to adapt the CIR [13]. The *in silico* results were promising, but a clinical trial is needed for confirmation.

To address the challenges related to the inter- and intra-patient variabilities and achieve personalisation of the insulin treatment, reinforcement learning (RL) has been introduced [14]. RL is a branch of machine learning (ML) that allows systems to develop self-learning abilities and thus to interact within uncertain environments. Moore *et al.* [15] introduced RL for optimal control of propofol-induced hypnosis. In a subsequent study in healthy human volunteers, the RL agent demonstrated clinically appropriate performance [16], [17]. In [18], a model-free RL-based control algorithm was implemented and validated *in silico* for its ability to deal with inter- and intra-patient variability and environmental uncertainties. The diabetic population in this study wore CGM and was treated with an insulin pump. The algorithm updated the BR and CIR each day on the basis of the patients' glucose level the day before. The algorithm's tuning was personalised and automatically based on the transfer entropy (TE) from insulin to glucose signals [19].

The research presented here is a continuation of our previous studies [14], [18] and targets the entire insulin pump population of adults with diabetes Type 1, independently of the technology used for glucose monitoring. The algorithm allows daily adjustment of the insulin infusion profile to compensate for fluctuation in the patient's glucose level. Information from SMBG or CGM provides input to the algorithm, which outputs the daily BR and three CIRs per day – one value for each of the three main meals. The self-learning approach is adaptable and personalises the daily insulin values to ensure glucose control, despite the inter- and intra-patient variabilities. The approach is data-driven, real-time and of low computational cost. To validate the newly introduced algorithm, an FDA-approved diabetes simulator was used.

## II. METHODOLOGY

The structure of the proposed dual mode adaptive basal-bolus advisor (ABBA) - along with its inputs and outputs - is illustrated in Fig. 1. Each day, ABBA provides one constant BR and three CIRs. Laimer *et al.* [20] analysed the BR profiles of 3118 female and 2427 male patients, and concluded that BR profiles with higher variability are associated with an increased frequency of acute complications in adults with diabetes Type 1. The study considered the dawn phenomenon as a factor influencing intra-day variability in insulin sensitivity, while the effect of intensive physical exercises was not taken into account. Furthermore, Bouchonville *et al.* [21] found that - for patients with insulin pumps - changing the basal rate in the early morning could not reduce the influence of the dawn phenomenon, but increased the risk of hypoglycaemia. Thus, in this study, the BR was considered as constant within a single day. To address the intra-day variation in insulin sensitivity (SI) during different meal timings, three different CIRs for breakfast, lunch and dinner were considered (Fig. 2).

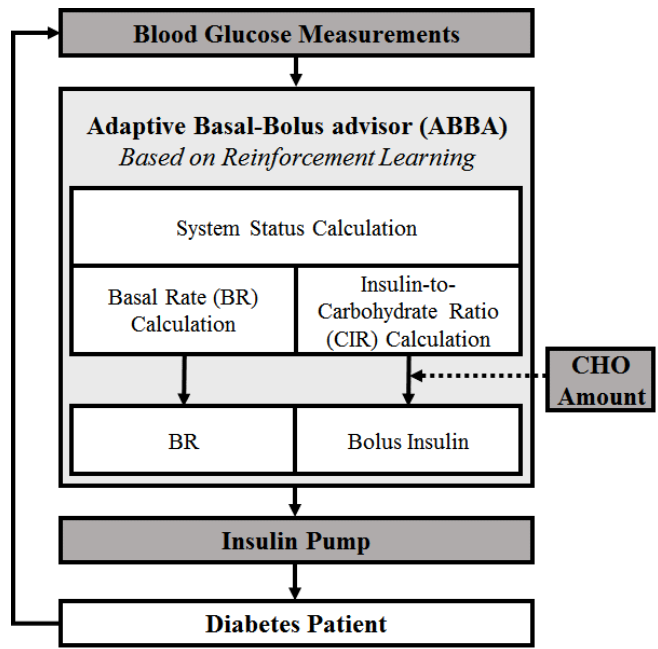

Fig. 1. Structure of ABBA with inputs and outputs.

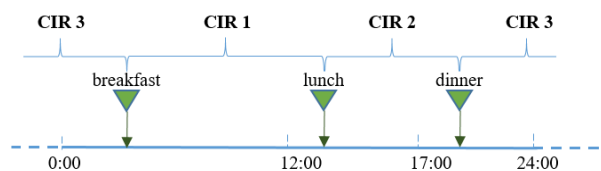

Fig. 2. Main meals and the corresponding CIRs.

ABBA employs the Actor-Critic (AC) method, a branch of RL, for updating BR and CIRs. The parameters of the actor-only method are directly estimated by simulation, and are updated in the direction of improvement. Critic-only methods rely exclusively on approximation of value function and aim to learn an approximate solution to the Bellman equation, which will then hopefully prescribe a near-optimal policy [22]. The AC method was selected because it combines the strong points of the actor-only and critic-only methods. In comparison with the critic-only method, for which convergence is guaranteed in limited settings, the AC method may converge in wider settings. On the other hand, it can achieve more rapid convergence than actor-only methods. In the next section, we will give a brief introduction to the AC method.

## A. Actor-Critic (AC) method

The critic uses an approximation architecture and simulation to learn a value function, which is then used to update the actor's policy parameters in the direction of performance improvement [22]. The AC method was introduced to minimise the average cost function $\bar{\alpha}$ as defined by:

$$\bar{\alpha}(\theta) = \sum_{x \in X, u \in U} c(x, u) \eta_\theta(x, u), \qquad (1)$$

where $c(x, u)$ is local cost, $\eta_\theta(x, u)$ is the stationary probability of the Markov chain $\{X_k, U_k\}$, $x$ is the state, and $u$ is control action.

The critic agent evaluates the current control policy through the approximation of the long-term expected cost. The critic provides temporal difference (TD) error to the Actor for policy optimisation. The value function is defined by

$$V_\theta(x) = E[\sum_{k=0}^{\infty} \gamma^k c(x_k, u_k) | x_0 = x], \qquad (2)$$

which can be formalised as:

$$V_\theta(x) = c(x_k, u_k) + \gamma V_\theta(y), \qquad (3)$$

where $\gamma$ is a discount factor in the range $\gamma \in (0,1)$, $y$ is the next state $y = x_{k+1}$.

Linear approximation was used for the parameterised function:

$$\tilde{V}_\theta^w(x) = \sum_{i=1}^{k} w^i g_\theta^i(x) = w^T g_\theta(x), \qquad (4)$$

where $w^T$ is the transpose of the parameter vector $w$ and $g_\theta(x)$ is a vector of basis function. The estimation of TD error $d$ can then be defined as:

$$d = c(x, u) + \gamma \tilde{V}_\theta^w(y) - \tilde{V}_\theta^w(x). \qquad (5)$$

The parameter vector $w$ is updated with the TD error:

$$w_{k+1} = w_k + a_k d_k z_k, \qquad (6)$$

where $a_k$ is a positive non-increasing learning rate sequence and $z_k$ is the eligibility vector updated according to:

$$z_{k+1} = \lambda z_k + g_\theta(x_{k+1}). \qquad (7)$$

The update for approximation of the action-value function follows a similar approach.

The actor agent aims to optimise the control policy in order to achieve the final goal of the AC method, i.e. to minimise the average cost function $\bar{\alpha}$ shown in equation (1). The policy gradient method is employed for this purpose:

$$\theta_{k+1} = \theta_k - \beta_k \nabla_\theta \bar{\alpha}(\theta), \qquad (8)$$

where $\beta_k$ is learning rate and $\nabla_\theta \bar{\alpha}(\theta)$ the gradient of $\bar{\alpha}(\theta)$ with respect to the policy parameter vector $\theta$, as calculated by:

$$\nabla_\theta \bar{\alpha}(\theta) = \sum_{x,u} \eta_\theta(x, u) d_t \psi_\theta(x, u), \qquad (9)$$

where $d_t$ is the TD error at time $t$ and $\psi_\theta(x, u)$ is the basis function for the action-value function.

## B. SMBG version of ABBA

### 1) SMBG measurements as system inputs

The SMBG version of ABBA (ABBA$_{SMBG}$) was designed to determine the "system status" (features) using four blood glucose measurements: before breakfast, lunch, dinner, and bedtime. This feature vector was used to update the control policy.

Specifically, a day's glycaemic profile was described by two types of features, $F_{hyper}$ and $F_{hypo}$, which were related to the system's hyperglycaemic and hypoglycaemic status, respectively. For calculation of these two features, we used the lower and upper border of tight target range, i.e. $G_L$=90 mg/dL and $G_H$=150 mg/dL, as thresholds:

$$F_{k\_hyper}^{BR} = \frac{1}{n_{k\_hyper}} \sum (M_{k\_hyper} - G_H) \qquad (10)$$

$$F_{k\_hypo}^{BR} = \frac{1}{n_{k\_hypo}} \sum (G_L - M_{k\_hypo}), \qquad (11)$$

where $F_{k\_hyper}^{BR}$ and $F_{k\_hypo}^{BR}$ are the features in the $k$-th day for the updated BR for the next day, $M_{k\_hyper}$ and $M_{k\_hypo}$ are the SMBG values that are above $G_H$ and below $G_L$, respectively. $n_{k\_hyper}$ and $n_{k\_hypo}$ are the numbers of $M_{k\_hyper}$ and $M_{k\_hypo}$ in the $k$-th day. If $n_{k\_hyper}$ or $n_{k\_hypo}$ is 0, the corresponding feature will have the value 0.

The feature calculation of $F_{k\_hyper}^{CIRi}$ and $F_{k\_hypo}^{CIRi}$ for the three CIRs follow a similar approach as for BR. The $i$ in the superscript enumerates the corresponding CIRs (1: breakfast, 2: lunch or 3: dinner). Using different features for BR and CIRs, it is possible to update the BR and CIRs in a relatively independent manner. In previous work [18], the same features were used for both BR and CIR, and the basal and bolus insulin always changed simultaneously in the same direction, i.e. the algorithm always offers increased basal insulin along with increased bolus insulin and vice versa. In order to overcome this limitation, we introduced three CIRs with different features and different update rules for BR and CIRs, as explained in the next section.

Both for BR and CIRs, the features were normalised into the range [0, 1], and the normalised feature could be presented in vector format:

$$F_k = (F_{k\_hyper}, F_{k\_hypo})^T. \qquad (12)$$

With these features, a local cost $c$ could be defined as:

$$c_k = a_{hyper} F_{k\_hyper} + a_{hypo} F_{k\_hypo}, \qquad (13)$$

where $a_{hyper} = 1$ and $a_{hypo} = 10$ are the scale parameters for weighting the hyperglycemic and hypoglycemic features. The critic part of ABBA could be updated as described in [18].

*2) Update process*

The update of BR and CIR from day $k$-1 to day $k$ considers the values of day $k$-1:

$$BR_k = BR_{k-1} + P_k^{BR}BR_{k-1} \tag{14}$$
$$CIR_{i_k} = CIR_{i_{k-1}} + P_{i_k}^{CIR}CIR_{i_{k-1}}, \tag{15}$$

where $P_k^{BR}$ and $P_{i_k}^{CIR}$ are the control actions for update BR and for CIRs in the $k$-th day. The subscript $i$ in (15) defines the type of the meal for which the CIR is applied (1 for breakfast, 2 for lunch and 3 for dinner).

To simplify the description of the equations, we introduced a new variable $AP$ to represent $BR_k$ and $CIR_{i_k}$ from equations (14) and (15), and named the final control action as $P_e$ to replace both $P_k^{BR}$ and $P_{i_k}^{CIR}$. Thus, equations (14) and (15) can be summarised as:

$$AP_{new} = AP_{old} + P_e AP_{old}, \tag{16}$$

where $AP_{new}$ is the value of BR and CIRs on day $k$, while $AP_{old}$ is the value on day $k$-1. In order to achieve a smooth update of BR and CIRs, we introduced a fusion value of $AP_{new}$ and $AP_{old}$:

$$AP_{fusion} = mAP_{old} + (1-m)AP_{new}. \tag{17}$$

The value of $m$ was experimentally chosen to be 0.5. According to equations (16) and (17), the fused AP was defined as:

$$\begin{aligned} AP_{fusion} &= 0.5 \cdot AP_{old} + (1-0.5)(AP_{old} + P_e AP_{old}) \\ &= AP_{old} + 0.5 \cdot P_e AP_{old}. \end{aligned} \tag{18}$$

For the BR update, the final BR was identical to the fused BR value:

$$BR_{final} = BR_{fusion}. \tag{19}$$

In order to avoid simultaneous increase/decrease of basal insulin and bolus insulin, an additional rule was established for updating CIRs:

$$CIR_{i\_final} = lCIR_{i\_old} + (1-l)CIR_{i\_fusion}, \tag{20}$$

where $l$ is a switch parameter (0 or 1) that specifies whether the final CIR should be the same as the fused value or the previous value. The $l$ parameter is defined by the following equations:

$$l = \begin{cases} 1 & \text{if } BR_{final} > BR_{old} \text{ and } CIR_{i\_fusion} < CIR_{i\_old} \\ & \text{or } BR_{final} < BR_{old} \text{ and } CIR_{i\_fusion} > CIR_{i\_old} \\ 0 & \text{others} \end{cases} \tag{21}$$

A further constraint was considered to limit the maximum change from $AP_{old}$ to $AP_{final}$ within 5%.

As in [14], the $P_e$ in this work consists of three parts: the linear deterministic control action $P_a$, the supervisory control action $P_s$, and the exploratory part $N(0, \sigma)$, which could be presented as Gaussian noise with zero mean and standard deviation $\sigma$. $\sigma$ is calculated as follows:

$$\sigma = c_\sigma \|F_k\|^2, \tag{22}$$

where the coefficient $c_\sigma$ has value 0.05. The value of $\sigma$ depends on the performance of the controller in the previous iteration. If ABBA achieves an optimised policy, i.e. the feature $F_k \rightarrow 0$, the exploration for next iteration is reduced correspondingly.

The calculation of $P_e$ can be described as:

$$P_e = hP_a + (1-h)P_s + N(0, \sigma), \tag{23}$$

where $h = 0.5$ and is a weighting factor to balance the contribution of $P_a$ and $P_s$ to the final control action $P_e$. The sum of the first two terms in equation (23) could be named as $P_d$:

$$P_d = hP_a + (1-h)P_s \tag{24}$$

Both $P_a$ and $P_s$ are calculated on the basis of the features $F_k$. The linear deterministic control action $P_a$ is defined as the linear combination of the features and policy parameter vector $\theta$:

$$P_a = F_k^T \cdot \theta_k \tag{25}$$

In this work, the calculation of $P_s$ for CIR is similar to that described in [18], i.e.

$$P_{s\_CIRi} = \begin{cases} 0 & \text{if } F_{k\_hyper}^{CIRi} = F_{k\_hypo}^{CIRi} = 0 \\ -0.02\, F_{k\_hyper}^{CIRi} & \text{if } F_{k\_hyper}^{CIRi} > 0 \text{ and } F_{k\_hypo}^{CIRi} = 0 \\ +0.02 F_{k\_hypo}^{CIRi} & \text{if } F_{k\_hypo}^{CIRi} > 0 \end{cases} \tag{26}$$

where $i$ indicates the $i$-th CIR. $F_{k\_hyper}^{CIRi}$ and $F_{k\_hypo}^{CIRi}$ are the hyperglycaemic and hypoglycaemic features, respectively.

The calculation of $P_s$ for BR was modified by evaluating the values of the measurements in different glucose level ranges:

$$P_{s\_BR} = \begin{cases} -F_{k\_hypo}/8 & \text{if } Hyponumber > 0 \\ F_{k\_hyper}/30 & \text{if } N_1 \le 1 \text{ and } N_2 \ge 2 \\ -F_{k\_hypo}/30 & \text{if } N_1 \ge 2 \text{ and } N_2 \le 1 \\ 0 & \text{others} \end{cases}, \tag{27}$$

where $Hyponumber$ is the number of measurements which are below 70 mg/dL. $N_1$ is the number of measurements below 80 mg/dL, while $N_2$ is the number of measurements above 130 mg/dL. The variables $Hyponumber$, $N_1$ and $N_2$ represent an overall trend of glucose level of the previous day.

Finally, the policy parameter update was defined as:

$$\theta_{k+1} = \theta_k - \beta d_k \frac{P_e - P_d}{\sigma^2} \nabla_\theta P_d, \qquad (28)$$

where $\beta$ is the actor learning rate values 0.5, and $d_k$ is the TD error.

A one week initialisation phase was applied before the normal control phase of ABBA. During the initialisation phase, the patients used their regular treatment. A CGM device was used for collecting blood glucose measurements for initialisation. With the seven-day measurements, the control policy parameter is initialised with the TE method, as described in [19].

### C. CGM version of ABBA

The CGM version of ABBA (ABBA$_{CGM}$) follows a similar approach to that of the SMBG version for "system status" calculation and update of BR and CIRs. The calculation of $P_s$ was modified as below:

$$P_s = \begin{cases} 0 & if \ F_{k\_hypo} = 0 \\ \mp 0.1 F_{k\_hypo} & if \ F_{k\_hypo} > 0 \ and \ F_{k\_hyper} = 0 \\ \mp 0.05 F_{k\_hypo} & if \ F_{k\_hypo} > 0 \ and \ F_{k\_hyper} > 0 \end{cases} \quad (29)$$

The upper sign in (29) refers to the calculation for BR and the lower sign to CIR. Like the SMBG version of ABBA, a one week initialisation phase with the TE method was applied before the control phase.

## III. EXPERIMENTAL PROTOCOL

### A. Simulation Environment

The two versions of ABBA were evaluated using the FDA-accepted adult population (100 virtual subjects) with the UVA/Padova T1DM simulator [23][24]. The simulator's default pump was selected for both CGM and SMBG versions of ABBA. As regards the glucose monitoring devices, ABBA$_{CGM}$ used Dexcom50 CGM with a sampling time of 5 minutes. This CGM was also used during the initialisation of both versions of ABBA, while ABBA$_{SMBG}$ used the default SMBG device during the operational period.

In the *in silico* environment, the system defines the type of meal based on the meal time. In fact, the user announces the meal by providing the CHO content of the upcoming meal. In our experiment, no bolus insulin was considered for bedtime snacks.

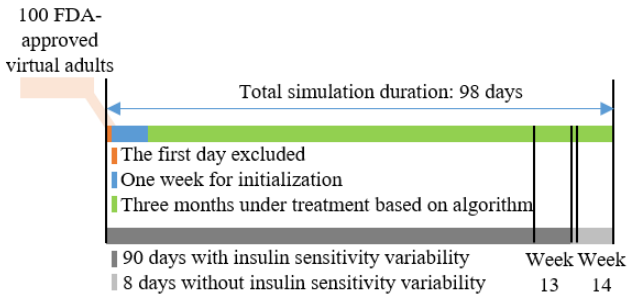

Fig. 3. Illustration of *in silico* evaluation settings.

### B. Experimental Protocol

The proposed approach was tested *in silico* on 100 simulated adults of the FDA accepted UvA/Padova Simulator using a number of scenarios emulating equivalent number of *in silico* clinical trials. Each trial lasted for 98 days (3 months and 1 week), excluding day 1 (no insulin on board is considered for day 1). Each patent's data from day 2 (D2) to D8 was used to initialise the control policy parameters. An initialisation period of seven days was chosen to include the weekly cycle of insulin sensitivity change, since the patient may have different behaviours over weekdays and weekends. During the initialisation period, the BR and CIR provided by the simulator were used to simulate standard treatment (ST). From D9 to D98, a period of 3 months, the ABBA was active. Dawn phenomenon and inter-day SI variability were considered until D90, while fixed SI was employed during the last 8 days (D91-D98). The last two weeks (D84-D90: Week 13 (W13) and D92-D98: Week 14) were used to evaluate the performance of ABBA against the OL period (D2-D8: W1). The D91 was excluded from evaluation since it was the transition day from *with SI variability* to *without SI variability*. The experimental protocol is illustrated in Fig. 3.

#### 1) Inter-day Variability of Insulin Sensitivity and Dawn Phenomenon

The inter-day variability of SI was simulated with a uniformly distributed variability of ±25%. The intraday variability usually caused by the dawn phenomenon was also considered. Dawn phenomenon, originally described in [25], refers to periodic episodes of hyperglycaemia occurring in the early morning hours before and after breakfast [26]. In that work, SI dropped every day between 04:00 and 08:00 to 50% of its nominal value, and SI ramped up or down within a timeframe of 30 minutes.

#### 2) Meal Protocol

Four meals of specific CHO content were considered for each day during the *in silico* trials: breakfast at 07:00 (50 g), lunch at 12:00 (60 g), dinner at 18:30 (80 g) and bedtime snack at 23:00 (15 g). Meal variability was introduced by considering a meal size variability of ±10 g for main meals and ±5 g for the bedtime snack and a meal-time variability of ±15 minutes. Furthermore, an uncertainty of ±50% in the CHO estimation was introduced. Both variabilities and uncertainties followed uniform distributions. Furthermore, the random skip of two main meals per week was considered (the corresponding insulin bolus was also skipped).

#### 3) Glucose measurements

In the case of ABBA$_{SMBG}$, the four glucose measurements of the previous day were used to update the BR and CIRs. The three pre-meal measurements were considered 20 minutes before the main meals, while the bedtime measurement took place at 23:00h. No pre- and postprandial measurements were taken for snacks and no bolus insulin infusion for bedtime snacks was required. All the measurements were used to estimate the "system status" (features) for BR, while for the case of CIRs only the measurements corresponding to the respective time window, i.e. the measurements till the next CHO announcement of main meal, were taken into consideration.

TABLE I
GLUCOSE LEVELS (MEAN ± STANDARD DEVIATION)

| | **D02-D08: Week 1 (standard treatment)** | | | | | **D84-D90: Week 13 (with SI variability)** | | | | | **D92-D98: Week 14 (without SI variability)** | | | | |
|---|---|---|---|---|---|---|---|---|---|---|---|---|---|---|---|
| | % in target range | % in Hypo | % in Severe Hypo | % in Hyper | % in Severe Hyper | % in target range | % in Hypo | % in Severe Hypo | % in Hyper | % in Severe Hyper | % in target range | % in Hypo | % in Severe Hypo | % in Hyper | % in Severe Hyper |
| *S1* | | | | | | 85.9±12.9 | 1.0±1.0 | 0.3±0.8 | 12.8±12.1 | 0.0±0.0 | 89.8±7.9 | 0.3±0.9 | 0.1±0.5 | 9.8±7.5 | 0±0.1 |
| *S2* | 89.9±8.7 | 2.5±3.0 | 1.5±3.3 | 6.1±8.2 | 0.0±0.1 | 84.2±12.8 | 0.5±0.8 | 0.2±0.6 | 15.2±12.4 | 0.0±0.0 | 88.5±8.8 | 0.2±0.6 | 0.1±0.4 | 11.2±8.4 | 0.1±0.4 |
| *S3* | | | | | | 84.8±12.6 | 0.4±0.7 | 0.1±0.4 | 14.7±12.3 | 0.0±0.0 | 88.7±8.7 | 0.2±0.7 | 0.1±0.4 | 11.0±8.5 | 0.1±0.4 |
| *S4* | - | | | | | 78.4±15.2 | 0.1±0.3 | 0.0±0.1 | 21.5±15.0 | 0.1±0.5 | 88.7±9.3 | 0.3±0.7 | 0.1±0.3 | 11.0±9.1 | 0.0±0.2 |

In the case of ABBA$_{CGM}$, all the CGM measurements of the previous day were used to estimate the "system status" and update the BR, while for the new CIR$_i$, all the CGM measurements for CIR$_i$ of the previous day, i.e. between previous day's CHO announcement for CIR$_i$ and its next CHO announcement, were considered.

Whenever the last measurement of the day was available (announced by the patient in the case of ABBA$_{SMBG}$ or at midnight in the case of ABBA$_{CGM}$), the new flat BR was estimated and activated to be used for the entire day. For the intraday CIRs, whenever a new meal was announced, the current CIR was deactivated and the CIR for the upcoming meal was estimated and activated. The update process was visualised in Fig. 4.

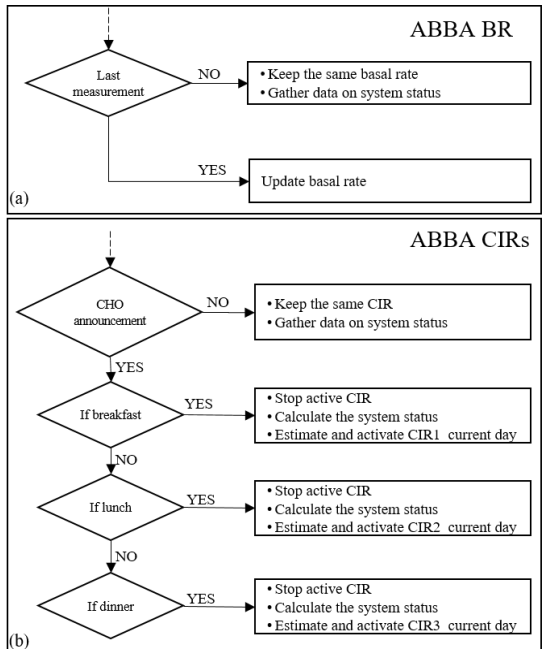

Fig. 4. Update process of BR and CIRs for one day.

*4) Scenarios*

Four *in silico* scenarios were considered:

- *Scenario 1 (S1)*: Combined use of CGM, ABBA$_{CGM}$ and insulin pump;
- *Scenario 2 (S2):* CGM for initialisation phase, SMBG, ABBA$_{SMBG}$ and insulin pump;
- *Scenario 3 (S3)*: Identical to S2 + uncertainty on SMBG measurement time;
- Scenario 4 (*S4*): Identical to S3 + skip of main meals.

In order to mimic real life situations, an uncertainty of ±10 min on standard glucose measurement time was considered in Scenarios 3 and 4.

*5) Evaluation metrics*

To evaluate and comparatively assess the performance of each approach, the following widely used metrics were implemented: percentage time in glucose target range [70,180] mg/dl; percentage time in hypoglycaemia [50 70) mg/dl; percentage time in severe hypoglycaemia <50 mg/dl; percentage time in hyperglycaemia (180, 300] mg/dl; and percentage time in severe hyperglycaemia >300 mg/dl. In addition, the low blood glycaemic index (i.e. risk of hypoglycaemia; LBGI), high blood glycaemic index (i.e. risk of hyperglycaemia; HBGI), the mean amplitude of glycaemic excursion (MAGE), and the total daily insulin intake (TDI) in units of insulin were estimated.

## IV. RESULTS AND DISCUSSION

Table I presents the *in silico* results observed in the tested scenarios. The ABBA$_{SMBG}$ versions (*S2* and *S3*) achieved comparable performance to ABBA$_{CGM}$ (*S1*), although only few SMBG measurements per day were available. In *S2* and *S3*, the number of hypoglycaemic events was further reduced. The percentages in target range were slightly decreased, mainly due to the increase in hyperglycaemic events. This increase was anticipated, since ABBA was designed to give high priority to hypoglycaemia, the more dangerous metabolic state. Furthermore, as expected, the percentages in hyper-and hypoglycaemic ranges during W14 (evaluation phase without SI variability) were lower than during W13 (evaluation phase with SI variability).

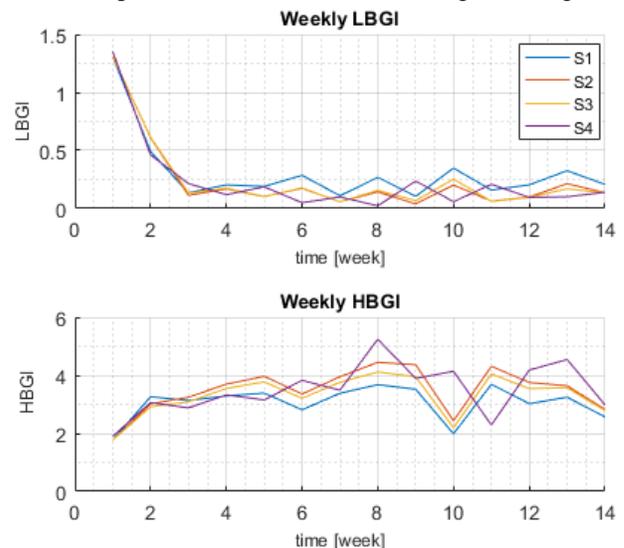

Fig. 5. Weekly LBGI and HBGI trends in 98-day trial.

The comparison of W1 (standard treatment) to W13 indicates that both ABBA$_{CGM}$ and ABBA$_{SMBG}$ significantly decreased the percentage of time in hypo- and severe hypoglycaemic ranges (Wilcoxon tests, $p <0.05$), while the respective percentages for hyperglycaemia were increased. The weekly LBGI and HBGI [27] are illustrated in Fig. 5. In all scenarios, the LBGI value was decreased from low range (1.1 - 2.5) in W1 to minimal range ($< 1.1$) in W13, while HBGI remained in minimal range ($<5$). After W3 (2$^{nd}$ week of ABBA), both LBGI and HBGI converged. During this two-week transition phase, ABBA progressively decreases the value of LBGI, and keeps HBGI within minimal range. The fact that HBGI was not increased over the trial period shows that the increase in the hyperglycaemias in all scenarios remained within the acceptable range. Furthermore, LBGI in the case of ABBA$_{SMBG}$ was lower than in the case of ABBA$_{CGM}$, while the opposite was observed for HBGI.

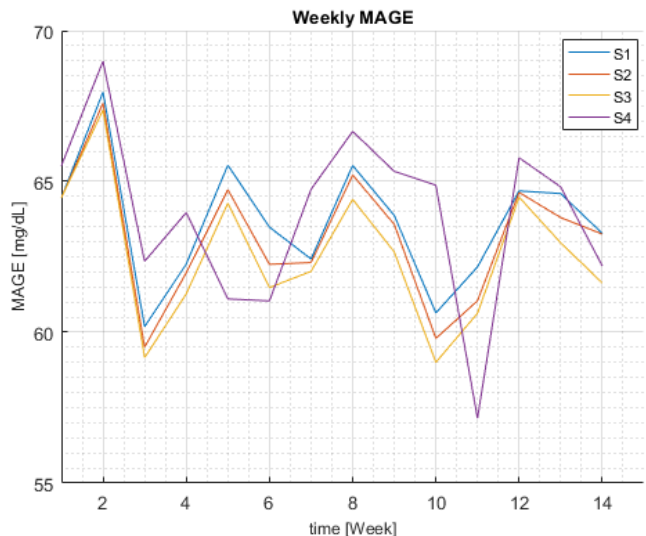

Fig. 6. Weekly mean amplitude of glycaemic excursion (MAGE) of the 100 subjects

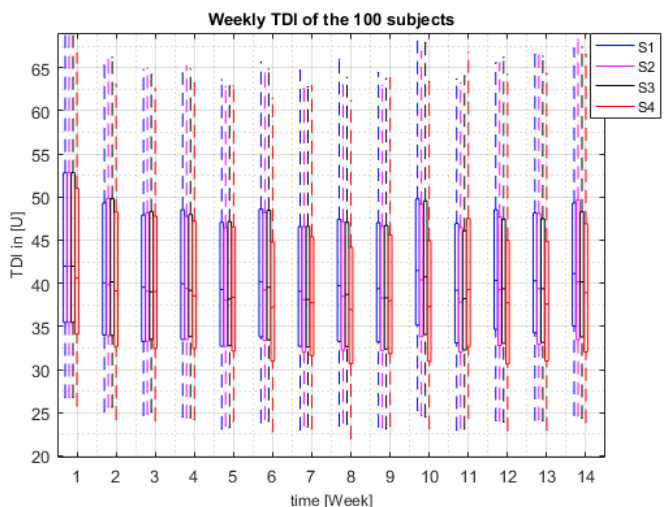

Fig. 7. Weekly mean total daily insulin (TDI) of the 100 subjects.

Fig. 6 presents the weekly mean value of mean amplitude of glycaemic excursions (MAGE) among the 100 subjects. The MAGE value indicates diabetic instability; a small MAGE value indicates more stable blood glucose concentration [28]. In comparison to $S1$, both $S2$ and $S3$ slightly decreased the MAGE value. The MAGE value of $S4$ shows that blood glucose regulation in this scenario is not as stable as in the other scenarios, since two meals per week were randomly skipped. The box plot in Fig. 7 shows the distribution of weekly mean total daily insulin (TDI) of the 100 subjects during the 98-day trial. In each week, both $S2$ and $S3$ of ABBA$_{SMBG}$ had similar median values and distributions to ABBA$_{CGM}$ ($S1$). As for S4, since two meals per week along with the corresponding bolus insulin were randomly skipped, the TDI was clearly lower than in the other scenarios.

## V. IMPLEMENTATION

Both versions of ABBA are easily applied to diabetic patients treated with insulin pumps. During the first seven operation days, ABBA provides the patient's standard treatment and, in parallel, collects the CGM and insulin pump data. For the case of SMBG users, the CGM can be provided by the attending physician. At the end of this period, the algorithm automatically estimates the TE and initialises the policy parameters, and is then ready to provide personalised insulin treatment with daily adaptation of BR and three CIRs based on either CGM or SMBG data. The patient can decide whether to accept or reject the suggested change. If the patient believes the ABBA suggested change exceeds his own estimation and decides to reject the value, he can choose to use the previous value or manually enter a new value for BR or CIRs.

ABBA$_{CGM}$ and ABBA$_{SMBG}$ are implemented in an Android platform. On this Android Platform, Debiotech's JewelPUMP application allows the patient to monitor and control the insulin pump. A communication protocol between ABBA and the JewelPUMP applications was defined and implemented, on the basis of standard Android Inter-Process Communication (IPC) mechanisms that allows communication between activities, as depicted in Fig. 8. In particular, the communication mechanism allows JewelPUMP to send messages to ABBA when a) there are basal profile changes, b) or bolus infusion or c) SMBG measurements are performed. The communication mechanism also enables ABBA to inform JewelPUMP about BR or bolus updates. When the patient announces a meal or the last BG measurement of the day, data synchronisation is performed, in order to ensure that all messages were properly sent from JewelPUMP to ABBA, and to send any that were not.

The implemented communication mechanism:

- Implements Inter-Process Communication (IPC) between JewelPUMP and ABBA.
- Allows JewelPUMP to send messages to ABBA.
- Ensures these messages are properly received by ABBA.

- Enables History Synchronisation.
- Allows synchronised bidirectional communication between JewelPUMP and ABBA.

With this communication protocol in place, the JewelPUMP Application is able to send and receive information to and from the ABBA application, in order to propose these personalised CIR values and basal rates to the patient, and subsequently to apply these values when controlling the infusion through the insulin patch pump.

The aforementioned implementation was conducted and tested on JewelCOM, an Android 4.4.4 based mobile platform. However, the ABBA application could be installed on smartphones with other Android version as well. In that case, compatibility issues need to be considered.

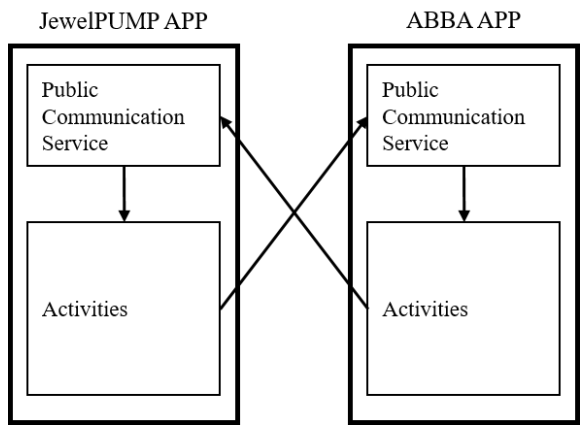

Fig. 8. Communication between JewelPUMP and ABBA applications.

## VI. CONCLUSIONS

An RL-based adaptive basal-bolus advisor, ABBA, is proposed. The advisor aims to minimise the risk of hypoglycaemia by providing personalised suggestions on daily BR and bolus dose on the basis of glucose measurements from either CGM or SMBG devices. The proposed approach was evaluated *in silico* on 100 adults from the FDA-accepted UVa/Padova Simulator under a number of challenging scenarios. A wide variety of different scenarios have been published, with different meal schemes and variability in insulin sensitivity. These are often combined with disparate variabilities and uncertainties. Therefore, it is not straightforward to compare performance in the present study with other publications. To this end, we considered four scenarios to evaluate both ABBA$_{SMBG}$ and ABBA$_{CGM}$, which were more challenging than those included in our previous research in the field. These scenarios consisted of complex meal protocols, including uncertainties about the size of the announced meal CHO and variabilities in meal announcement times, inter- and intra-day variabilities in insulin sensitivity and dawn phenomenon, as well as uncertainties about the time of SMBG glucose measurements. The performance of ABBA$_{SMBG}$ and ABBA$_{CGM}$ converged after two weeks of operation, while, during the transition phase, both versions of ABBA progressively achieved better glucose control in terms of LBGI. The results indicate that - independent of the technology used for glucose measurement - the proposed RL approach is able to i) learn the patient's characteristics and ii) provide personalised suggestions on insulin treatment. The insulin suggestions virtually eliminated hypoglycaemias and maintained glucose in the target range most of the studied time, even in the case of extreme scenarios with uncertainties, variabilities, and skipped main meals.

Furthermore, the proposed approach relies on the standard medical treatment as starting point, is easily applied, and the SMBG version implements the NICE guidelines with respect to the minimum number of fasting glucose measurements per day. The two versions have already been integrated on Android smartphones that are able to communicate wirelessly with a patch pump.

The next step is to conduct a feasibility study within the framework of a pilot clinical trial to confirm the *in silico* results.